\documentclass[conference]{IEEEtran}

\usepackage{amsmath,amsfonts,amssymb}
\usepackage[utf8]{inputenc}
\usepackage{graphicx}
\usepackage{amsthm}

\usepackage{algorithmicx}
\usepackage{algorithm}
\usepackage{algpseudocode}
\algnewcommand{\IIf}[1]{\State\algorithmicif\ #1\ \algorithmicthen}
\algnewcommand{\EndIIf}{\unskip\ \algorithmicend\ \algorithmicif}

\usepackage{calc}   
\usepackage{multirow}

\makeatletter
\def\textSq#1{%
\begingroup
\setlength{\fboxsep}{0.3ex}
\setbox1=\hbox{#1}
\setlength{\@tempdima}{\maxof{\wd1}{\ht1+\dp1}}
\setlength{\@tempdimb}{(\@tempdima-\ht1+\dp1)/2}
\raise-\@tempdimb\hbox{\fbox{\vbox to \@tempdima{%
  \vfil\hbox to \@tempdima{\hfil\copy1\hfil}\vfil}}}%
\endgroup%
}
\makeatother

\newtheorem{theorem}{Theorem}
\newtheorem{definition}[theorem]{Definition}
\newtheorem{proposition}[theorem]{Proposition}

\newcommand{\pibar}{p^{\text{\sc ga}}_{i|}}
\newcommand{\ppibar}{{p'}^{\text{\sc ga}}_{i|}}
\newcommand{\G}{\mathbf{G}} 
\newcommand{\B}{\mathbf{B}}
\DeclareMathOperator{\atanh}{atanh}

\begin{document}

\title{On Puncturing Strategies for Polar Codes}

\author{\IEEEauthorblockN{Ludovic Chandesris\IEEEauthorrefmark{2}\IEEEauthorrefmark{3},
Valentin Savin\IEEEauthorrefmark{2}, David Declercq\IEEEauthorrefmark{3}  \\ {ludovic.chandesris@cea.fr,  valentin.savin@cea.fr,  declercq@ensea.fr}}
\IEEEauthorrefmark{2}CEA-LETI, Minatec campus, Grenoble, France \quad
\IEEEauthorrefmark{3}ETIS, ENSEA/UCP/CNRS, Cergy-Pontoise, France}

\maketitle

\begin{abstract}
This paper introduces a class of specific puncturing patterns, called symmetric puncturing patterns, which can be characterized and generated from the rows of the generator matrix $G_N$. They are first shown to be non-equivalent, then a low-complexity method to generate symmetric puncturing patterns is proposed, which performs a search tree algorithm with limited depth, over the rows of $G_N$. Symmetric  patterns are further optimized by density evolution, and shown to yield better performance than state-of-the-art rate compatible code constructions, relying on either puncturing or shortening techniques.  
%
\end{abstract}


\section{Introduction}\label{sec:num1}

Polar codes are a recently discovered family of error correcting codes \cite{arikan2009channel}, known to achieve the capacity of any binary-input memoryless output-symmetric channel. Their construction relies on a specific recursive encoding procedure that synthesizes a set of $N$ virtual channels from $N$ instances of the transmission channel, where $N$ denotes the code-length. The recursive encoding procedure is reversed at the receiver end, by applying a Successive Cancellation (SC) decoder. The asymptotic effectiveness of the SC decoder derives from the fact that the synthesized channels tend to become either noiseless or completely noisy, as the code-length goes to infinity, phenomenon which is known as ``channel polarization''.

A Polar codeword $x_0^{N-1} \in \{0, 1\}^N$ is obtained by applying a linear transformation on a {\em data vector} $u_0^{N-1} \in \{0, 1\}^N$ that contains both {\em information} and {\em frozen bits} ({\em i.e.}, bits whose values are known at both the transmitter and the receiver). The positions of  information and frozen bits are specified by a binary {\em information pattern} $I\subset\{0,1\}^N$, with $1$s corresponding to information bits, and $0$s corresponding to frozen bits. The linear transformation is defined by the square matrix $\G_N$, of size $N\times N$, defined as the $n$-th Kronecker product of the {\em kernel matrix} $\G_2$ \cite{arikan2009channel}: 
\begin{equation}
{G}_2=\begin{bmatrix}
   1 & 0 \\
   1 & 1
\end{bmatrix}
\end{equation}
Hence, $N=2^n$ and $x_0^{N-1} = u_0^{N-1}\cdot \G_N$, with $\G_N = \G_2^{\otimes n}$. However, in practical communication systems the length of the coded block may not be a power of $2$. In this case, punctured Polar codes have to be considered, in which only a number $N-N_p$ of the encoded bits $x_0^{N-1}$ are transmitted over the channel. The positions of the $N_p$ punctured bits are indicated by a {\em puncturing pattern} 
$P \subset\{0,1\}^N$. The optimization of the puncturing pattern amounts to determining the $N_p$ positions yielding the punctured code with the best possible decoding performance. It is worth noticing that the optimal solution requires a joint optimization of both $P$ and $I$, since different puncturing patterns may lead to different information patterns. Hence, a brute-force search algorithm would require testing all the possible puncturing patterns, {\em i.e.} $\binom{N}{N_p}$ possibilities, and for each puncturing pattern finding the optimal pattern of information bits, {\em e.g.}, by using the density evolution technique, as explained in Section~\ref{sec:num2}. In practice, this is  not feasible even for relatively small values of $N$ and $N_p$.  The optimization problem is slightly different in case of communication systems employing retransmission techniques, such as Hybrid Automatic Repeat reQuest (HARQ) \cite{hong2015capacity, saber2015incremental,chen2013hybrid,niu2013beyond}, since the overall system performance depends on the performance of both the punctured and the unpunctured (mother) code, which are further constrained by a dependency relation, as they both have to use the same information pattern $I$.    

While the problem of finding the optimal puncturing pattern is still open, several puncturing techniques have been proposed in the literature, aimed at enhancing the punctured Polar code performance, by either making use of different properties of the punctured code  (in terms of minimal distance or exponent bound) \cite{niu2013beyond,shin2013design,eslami2011practical}, or relying on the density evolution analysis \cite{kim2015efficient,zhang2014puncturing}. 
However, as no efficient method has been proposed to evaluate the optimal solution for moderate to long code-length, the gap between any method and the optimality is unpredictable. 

In this paper we introduce a number of theoretical tools aimed at analyzing and classifying puncturing patterns. To this end, we first introduce {\em equivalent puncturing patterns}, and show they yield punctured codes with the same error correction performance. Then, we give a constructive solution to the problem of finding a unique representative for each equivalence class of puncturing patterns. Such a representative is referred to as a {\em primitive puncturing pattern}, and we further distinguish between primitive patterns that are {\em symmetric} from those that are not. The symmetry property means that $P$ is equal to the {\em erasure pattern} $E$ that is generated on the data vector $u_0^{N-1}$, corresponding to virtual channels with symmetric capacity equal to zero (see Section~\ref{subsec:punct_polar_codes}). Then, we show that symmetric patterns can be generated by the rows of the generator matrix $\G_N$, and propose an algorithm able to generate symmetric patterns of a given {\em order} (the order of a symmetric pattern is defined as the minimum number of rows of $\G_N$ that contribute to generating it). This makes possible to implement a brute-force optimization algorithm on symmetric patterns of predetermined maximum order. Finally, the performance of the optimized puncturing patterns is evaluated and compared with state of the art solutions. 

\section{Preliminaries}\label{sec:num2}
  
\subsection{Notation}\label{subsec:notation}
\begin{list}{$\bullet$}{\setlength{\leftmargin}{10pt}\setlength{\labelwidth}{5pt}}
\item {\em Vectors} are denoted by either $a_0^{N-1}=(a_0,\dots,a_{N-1})$ or $A=\left(A(0),\dots,A(N-1)\right)$. {\em Matrices} are denoted by  boldface uppercase letters, {\em e.g.}, $\mathbf{M}$. 
\item $B_N:\{0,\dots,N-1\}\rightarrow \{0,\dots,N-1\}$ denotes the {\em bit-reversal} permutation, and $\B_N$ the corresponding binary permutation matrix, where $\B_N(i,j)=1 \Leftrightarrow j=B_N(i)$.
\item For any permutation $\phi:\{0,\dots,N-1\}\rightarrow \{0,\dots,N-1\}$ and any vector $A=\left(A(0),\dots,A(N-1)\right)$, $\phi(A)$ is the vector defined by $\phi(A)(i) = A(\phi(i))$.
\item $|A|$ denotes the number of of $1$s of a binary vector $A$.
\item $A = \vee_{\ell=1}^{L} A_\ell$ denotes the {\em bit-wise {\sc or} operation} of a set of {\em binary vectors} $A_1,\dots, A_L$, where $A(i) = 1 \Leftrightarrow \exists \ell \in \{1,\dots, L\}, \text { such that }  A_\ell(i) = 1$. By a slight abuse of language, we shall also say that {\em  $A$ is the union of the vectors $A_1,\dots, A_L$.}
\item For two binary vectors $A, B$, we say that {\em $A$ is included in $B$}, and write $A \subseteq B$, if $A\vee B = B$. 
\item $A \leq_{lex} B$ denotes the lexicographic ordering of $A$ and $B$. Thus, assuming $A \ne B$, $A <_{lex} B \Leftrightarrow A(k)<B(k)$,  where  $k=\min\{i \mid A(i)\ne B(i)\}$
\end{list}

\subsection{Punctured Polar Codes}\label{subsec:punct_polar_codes}

Let $u_0^{N-1}$ be a data vector, $x_0^{N-1}=u_0^{N-1} \cdot \G_N $ the corresponding encoded vector\footnote{Note that in \cite{arikan2009channel},  encoding is defined by 
$x_0^{N-1}=u_0^{N-1} \cdot \widetilde{\G}_N$, where $\widetilde{\G}_N=\B_N \cdot \G_N$. Hence, $\widetilde{\G}_N$ differs from $\G_N$ by a column permutation, but the two constructions are known to be equivalent.}, and
$P \subset\{0,1\}^N$ a puncturing pattern (hence, $x_i$ is punctured if and only if $P(i)=1$). Assume that that transmission takes place over a Binary-Input Discrete Memoryless  Channel (B-DMC) $W$, with output alphabet $\mathcal{Y}$. We denote by $y \lbrack P \rbrack_0^{N-1}$ the punctured received sequence. For the sake of simplicity, we shall assume that $y \lbrack P \rbrack_0^{N-1}$ is a sequence of length $N$, with $y \lbrack P \rbrack_0^{N-1} \in (\mathcal{Y} \cup \{\star\})^N$ and $y \lbrack P \rbrack_i=\star \Leftrightarrow P(i)=1$ (the output alphabet is extended with the symbol $\star$ to account for punctured positions).

Let $I \subset\{0,1\}^N$ be an information pattern, with $|I| = K < N-N_p$. We denote by $(N,P,I)$ the corresponding punctured Polar code of dimension $K$ and length $N-N_p$.  Frozen bits are assumed to be set to zero, that is $u_i = 0$, for $I(i)=0$. Using similar notation to that of \cite{arikan2009channel}, we denote by $W \lbrack P \rbrack _N^{(i)}$ the virtual channel with input $u_i$ and output $(y \lbrack P \rbrack_0^{N-1},u_0^{i-1})$, and by $\mathcal{I}(W \lbrack P \rbrack _N^{(i)})$ the symmetric capacity of $W \lbrack P \rbrack _N^{(i)}$. For a given $P$, the optimal information set $I$ is given by the $K$ positions with highest $\mathcal{I}(W \lbrack P \rbrack _N^{(i)})$ values. 

The symmetric capacity $\mathcal{I}(W \lbrack P \rbrack _N^{(i)})$ can be determined from the distribution of the log-likelihood ratio (LLR) values $\gamma_i$, computed by the {\em genie-aided}\footnote{Following \cite{arikan2009channel}, we refer to such a decoder as {\em genie-aided}, since it assumes perfect knowledge of the previous data bits $u_0^{i-1}$.} SC decoder:
\begin{equation}
\gamma_i=\log \left(\frac{\Pr(u_i=0 \mid y \lbrack P \rbrack_0^{N-1}, u_0^{i-1})}{\Pr(u_i=1 \mid y \lbrack P \rbrack_0^{N-1}, u_0^{i-1})} \right),
\end{equation}
Let $\Gamma(W \lbrack P \rbrack _N^{(i)})$ denote the probability distribution function of $\gamma_i$. In practice,  $\Gamma(W \lbrack P \rbrack _N^{(i)})$ can be estimated, under the all-zero codeword assumption, by relying on the density evolution (DE) technique \cite{mori2009performance}. It further allows determining the {\em genie-aided} error probability: 
\begin{equation} 
\pibar=\Pr(\gamma_i<0)+\frac{1}{2} \Pr(\gamma_i=0),
\end{equation}
which can be used as an alternative metric for the choice of the information pattern $I$, by choosing by the $K$ positions with lowest $\pibar$ values.


\begin{definition}
The data bit erasure pattern is a binary vector ${E} \lbrack {P} \rbrack \in\{0,1\}^N$, such that ${E} \lbrack {P} \rbrack(i) = 1$ iff $\mathcal{I}(W \lbrack P \rbrack _N^{(i)})=0$.
\end{definition}
Assuming that $\mathcal{I}(W) > 0$, it can be shown that ${E} \lbrack {P}\rbrack$ does actually not depend on $W$, but only on the puncturing pattern ${P}$, as explained in Section~\ref{subsec:symmetric}. When no confusion is possible, we shall simply denote ${E} \lbrack { P} \rbrack$ by ${ E}$. It follows from the above definition that the erasure pattern indicates data bit positions $i \in\{ 0,\dots,N-1\}$ such that $\Gamma(W \lbrack P \rbrack _N^{(i)})$ is the Dirac distribution supported at zero. It is clear that the information pattern $I$ should be chosen such that $E \wedge I= \underline{0}$, where $\wedge$ denotes the bit-wise {\sc and} operation, and $\underline{0}$ the all-zero vector.
The following theorem is proved in \cite{zhang2014puncturing}. 
\begin{theorem}
$|{E} \lbrack { P} \rbrack|=|{P}|$, for any puncturing pattern $P$.
\end{theorem}

\subsection{Successive Cancellation Decoder}
The SC decoder takes advantage of the code construction, by estimating the value of a data bit $u_i$, denoted by $\hat{u}_i$, based on the previous estimates $\hat{u}_0^{i-1}$. To this end, the following LLR value is computed:
\begin{equation} 
\hat{\gamma}_i=\log \left(\frac{\Pr(u_i=0 \mid y \lbrack P \rbrack_0^{N-1}, \hat{u}_0^{i-1})}{\Pr(u_i=1 \mid y \lbrack P \rbrack_0^{N-1}, \hat{u}_0^{i-1})} \right),
\end{equation} 
then the data bit estimate is defined by $\hat{u}_i = \displaystyle\frac{1-\text{sign}(\hat{\gamma}_i)}{2}$, if $I(i)=1$, and $\hat{u}_i = u_i$, if $I(i)=0$ (note that $\text{sign}(0) = \pm1$ with equal probability).
Since the SC decoder has to rely on the previous estimates of the data bits, $\hat{\gamma}_i$ and ${\gamma}_i$ values  need not have the same probability distribution.  
However, in practice, the Word Error Rate (WER) performance of the SC decoder can be approximated by the following formula, which is known to be tight, especially in the high Signal to Noise Ratio (SNR) regime \cite{wu2014construction}:
\begin{equation}\label{eq:wer_approx}
\text{WER} \approx \text{WER}^{\text{\sc ga}} = 1-\prod_{i \in \mathcal{I}}(1-\pibar)
\end{equation}

\section{Equivalent and Primitive Puncturing Patterns}\label{sec:equiv_primitive_pattern}

\subsection{Equivalent Puncturing Patterns}

For $j=0,...,n-1$ and $k=0,2^{j+1},\dots,(2^{n-j-1}-1)2^{j+1}$, let $\phi_{k,j}:\{0,\dots,N-1\}\rightarrow\{0,\dots,N-1\}$ be the permutation swapping the elements within the blocks $[k,\dots, k+2^j-1]$ and $[k+2^j,\dots, k+2^{j+1}-1]$. Hence:
\begin{equation}
\phi_{k,j}(i) = \left\{\begin{array}{ll}
  i, & \text{if } i < k \text{ or } i \geq k+2^{j+1}\\
  i+2^j, & \text{if } k \leq i < k+2^j \\
  i-2^j, & \text{if } k+2^j \leq i < k+2^{j+1}
\end{array}\right.
\end{equation}
%
In the following, $\phi_{k,j}$ will be referred to as an {\em elementary permutation}. Clearly, $\phi_{k,j}\circ\phi_{k,j}$ is the identity permutation. 


\begin{definition}
Two puncturing patterns $P$ and $P'$ are said to be \textit{equivalent}, denoted by $P \approx P'$, if there exists a sequence of elementary permutations $\phi_{k_1,j_1},\dots, \phi_{k_t,j_t}$ that transform $B_N(P)$ into $B_N(P')$. Hence:
\begin{equation}\label{eq:equivalent_patterns}
P' = B_N(\phi_{k_t,j_t}(\cdots(\phi_{k_1,j_1}(B_N(P)))\cdots))
\end{equation}
\end{definition}

\begin{table}
  \centering
  \caption{Non-Equivalent Patterns with Same Erasure Set}
  \label{tab:table1}
  \renewcommand{\arraystretch}{1.25}
  \begin{tabular}{|@{\,}c@{\,}|r@{\,}c@{\,}c@{\,}c@{\,}c@{\,}c@{\,}c@{\,}c@{\,}c|}
     \hline
     punct.  & \multicolumn{9}{l|}{\ ${\cal I} = \{\mathcal{I}(W \lbrack P \rbrack _7^{(i)})\mid i=0,\dots,7\}$} \\
     pattern & \multicolumn{6}{l}{\ $p = \{\pibar \mid i=0,\dots,7\}$} & \multicolumn{3}{r|}{$[W = \text{BEC}(\varepsilon=0.5)]$} \\
    \hline\hline
    \multirow{2}{5mm}{\centering $P$} &  
    ${\cal I} =\{$  & $0,$ & $0,$ & $0,$ & $\mathbf{0.25},$ & $0,$ & $\mathbf{0.375},$ 
                    & $\mathbf{0.4375},$ & $\mathbf{0.9375}\,\}$ \\
    & $p = \{$ & $0.5,$ & $0.5,$ & $0.5,$ & $\mathbf{0.375},$ & $0.5,$ & $\mathbf{0.3125},$ 
                    & $\mathbf{0.2813},$ & $\mathbf{0.0313}\,\}$\\
    \hline
    \multirow{2}{5mm}{\centering $P'$} & 
    ${\cal I}'=\{$ & $0,$ & $0,$ & $0,$ & $\mathbf{0.25},$ & $0,$ & $\mathbf{0.250},$ 
                    & $\mathbf{0.5625},$ & $\mathbf{0.9375}\,\}$\\
    & $p' = \{ $ & $0.5,$ & $0.5,$ & $0.5,$ & $\mathbf{0.375},$ & $0.5,$ & $\mathbf{0.375},$ 
                & $\mathbf{0.2188},$ & $\mathbf{0.0313}\,\}$ \\
    \hline
    \multicolumn{10}{l}{$P = [1 1 1 0 1 0 0 0],\ P' = [1 1 0 1 1 0 0 0],\ E[P] = E[P'] = P$}
  \end{tabular}
\end{table}

\begin{proposition}\label{prop:equiv_pattern}
Let $P' \approx P$, $y[P]_0^{N-1} \in (\mathcal{Y} \cup \{\star\})^N$, and 
$y' [P']_0^{N-1}$ be the permuted version of $y[P]_0^{N-1}$, defined by the permutation $B_N\circ\phi_{k_t,j_t}\circ\cdots\circ\phi_{k_1,j_1}\circ B_N$, from Eq.~(\ref{eq:equivalent_patterns}). Let ${\gamma}_0^{N-1}$ and ${\gamma'}_0^{N-1}$ be the LLR values computed by the genie-aided SC decoder, under the all-zero codeword assumption, when submitted with $y[P]_0^{N-1}$ and $y'[P']_0^{N-1}$ inputs, respectively. Then ${\gamma}_0^{N-1} = {\gamma'}_0^{N-1}$.
\end{proposition}

\noindent\textit{Proof}: It is enough to prove the proposition for $t=1$, since $B_N\circ\phi_{k_t,j_t}\circ\cdots\circ\phi_{k_1,j_1}\circ B_N = \left( B_N\circ\phi_{k_t,j_t}\circ B_N\right)\circ\cdots\circ\left(B_N\circ\phi_{k_1,j_1}\circ B_N\right)$. Assuming $t =1$, the proposition is equivalent to the fact that the genie-aided SC decoder on $\widetilde{\G}_N=\B_N \cdot \G_N$ is invariant by any elementary permutation $\phi_{k,j}$ of $y[P]_0^{N-1}$. This can be proved by induction, starting from the observation that permuting the $y_0$ and $y_1$ values on a kernel block (see Fig.~\ref{fig:my_label2}) does not change the values of $\gamma_0$ and $\gamma_1$, since under the all-zero codeword assumption $\gamma_0 = 2 \atanh(\tanh(\ell_0/2)\tanh(\ell_1/2))$ and $\gamma_1 = \ell_0+\ell_1$, where $\ell_i = \log\left(\Pr(x_i=0\mid y_i)/\Pr(x_i=1\mid y_i)\right)$, $i=0,1$. The induction step follows from the fact that $\phi_{k,j}$ permutes the half-top and half-bottom outputs of a $\widetilde{\G}_{2^{j+1}}$ sub-block of $\widetilde{\G}_N$. 
$\mbox{ }$\hfill$\square$

\begin{theorem}
Let $P$ and $P'$ be two equivalent patterns. Then, the following properties hold: 
\begin{itemize}
\item[$(i)$] $\Gamma(W \lbrack P \rbrack _N^{(i)})=\Gamma(W \lbrack P' \rbrack _N^{(i)}), \ \forall\,W, \forall i = 0, \dots, N-1$
\item[$(ii)$] $\mathcal{I}(W \lbrack P \rbrack _N^{(i)})=\mathcal{I}(W \lbrack P' \rbrack _N^{(i)}), \ \forall\,W, \forall i = 0, \dots, N-1$
\item[$(iii)$] $E[P] = E[P']$
\end{itemize}
\end{theorem}

\noindent\textit{Proof}: $(i)$ follows from Proposition~\ref{prop:equiv_pattern}, $(ii)$ follows from $(i)$, and $(iii)$ from $(ii)$.\hfill$\square$

In particular, it follows that the genie-aided SC decoder has the same error probability ($\pibar, i=0,\dots,N-1$), and therefore the same $\text{WER}^{\text{\sc ga}}$ (Eq.~(\ref{eq:wer_approx})), irrespective of which of $P$ or $P'$ is being used. 

It is worth noticing that there exist puncturing patterns with $E[P] = E[P']$, but which are not equivalent. An example is provided in Table~\ref{tab:table1} for the Polar code of length $N=8$, where $P = [1, 1, 1, 0, 1, 0, 0, 0]$ and $P' = [1, 1, 0, 1, 1, 0, 0, 0]$. It can be seen that $E[P] = E[P'] = P$, corresponding to indexes $i=0,1,2,4$ with $\mathcal{I}(W \lbrack P \rbrack _N^{(i)})=\mathcal{I}(W \lbrack P' \rbrack _N^{(i)}) = 0$. However, $\mathcal{I}(W \lbrack P \rbrack _N^{(i)})\neq \mathcal{I}(W \lbrack P' \rbrack _N^{(i)})$ for $i=5,6$, hence $P$ and $P'$ cannot be equivalent\footnote{In \cite{kim2015efficient, zhang2014puncturing}, it is stated without proof that $E[P] = E[P']$ if and only if $\pibar = \ppibar,\ \forall i=0,\dots,N-1$. Our example also shows that the statement in \cite{kim2015efficient, zhang2014puncturing} is actually false.}. 
Table~\ref{tab:table1} also provides the error probability values $\pibar$. Using Eq.~(\ref{eq:wer_approx}), it can be verified that, depending on the number of information bits, better $\text{WER}^{\text{\sc ga}}$ performance is provided by $P'$ if $K=2$ (with information bits $\{u_6, u_7\}$), or $P$ if $K=3$ (with information bits $\{u_5, u_6, u_7\}$).

Considering equivalence classes of puncturing patterns is particularly relevant when brute-force search is used to find the puncturing pattern (together with the corresponding information set) providing the best $\text{WER}$ performance. In the following section we introduce a constructive method to determine a unique representative in each  equivalence class, which allows reducing the complexity of the brute-force search.  


\subsection{Primitive Pattern of an Equivalence Class}


For a puncturing pattern $P$, let $\phi(P)$ be the puncturing pattern defined by Algorithm~\ref{SearchPr}.  Algorithm~\ref{SearchPr} visits all the elementary permutations in a specific order, but an elementary permutation $\phi_{k,j}$ is performed only if $[P(k),\dots,P(k+2^j-1)] <_{lex} [P({k+2^j}),\dots,P({k+2^{j+1}-1})]$. Note that in such a case, $\phi_{k,j}$ actually permutes the above two blocks. We further define:
\begin{equation}
\Phi(P) = B_N(\phi(B_N(P))
\end{equation}

\begin{algorithm}[!tb]
\caption{Function $\phi$}
\label{SearchPr}
\begin{algorithmic}[1]
\Procedure{$\phi$}{$P$}
	\For {$j=0:1:n-1$}	
	    \For {$k=0:2^{j+1}:N-1$}
    	 	\If {$P_k^{k+2^j-1} <_{lex} P_{k+2^j}^{k+2^{j+1}-1} $}
    	 	    \State $P \leftarrow \phi_{k,j}(P)$ 
    	 	\EndIf
		\EndFor
	 \EndFor
	\State return $P$
\EndProcedure
\end{algorithmic}
\end{algorithm}

\begin{theorem}
The following properties hold:
\begin{itemize}
\item[$(i)$] $\Phi(P) \approx P$ 
\item[$(ii)$] $P \approx P' \Leftrightarrow \Phi(P) = \Phi(P')$
\item[$(iii)$] $\Phi(P) = \displaystyle\mathop{\text{\rm argmax}_{\text{lex}}}_{P' \approx P} B_N(P')$
\end{itemize}
\end{theorem}
From the  above theorem, it follows  that $\Phi(P)$ only depends on the equivalence class of $P$. In the following, we say that $P$ is a {\em primitive pattern} if  $\Phi(P)=P$. Hence, each equivalence class contains one and only one primitive pattern.  The proofs of the above and the following theorem will be provided in an extended version of this paper.

\begin{theorem}\label{theo:primitive_reccursion}
Let $P$ be a puncturing pattern, $\bar{P} = B_N(P)$, $\bar{P}_0 = [\bar{P}(0),\dots, \bar{P}(N/2-1)]$ and 
$\bar{P}_1 = [\bar{P}(N/2),\dots, \bar{P}(N-1)]$. Then  $P$ is primitive if and only if $B_{N/2}(\bar{P}_0)$ and $B_{N/2}(\bar{P}_1)$ are primitive, and $\bar{P}_1\leq_{lex}\bar{P}_0$. 
\end{theorem}

The above theorem states that primitive patterns can be determined recursively. However, the recursive construction method requires storing all of the primitive patterns of length $N$, in order to derive those of length $2N$, thus becoming unfeasible for large $N$ values. In Section~\ref{sec:symmetric_patterns}, we restrict to a specific class of puncturing patterns, referred to as {\em symmetric}. We show that any symmetric pattern is primitive, and propose a search-tree algorithm that allows generating symmetric patterns even for large values of $N$.

\section{Symmetric Puncturing Patterns}\label{sec:symmetric_patterns}

\subsection{Symmetric Patterns}\label{subsec:symmetric}

\begin{definition}
A puncturing pattern ${P}$ is said to be symmetric\footnote{Note that using Arikan's generator matrix $\widetilde{\G}_N=\B_N \cdot \G_N$, symmetric patterns correspond to patterns $P$, such that ${E} \lbrack {P} \rbrack=B_N({P})$. Such patterns have been considered in \cite{kim2015efficient}, but the link with the rows of $\G_N$ (Theorem~\ref{theo:erasure_pattern}) has not been not pointed out.} if ${E} \lbrack {P} \rbrack={P}$.
\end{definition}

\noindent The following theorem provides a simple characterization of symmetric patterns, as the union of a subset of rows of $\G_N$.


\begin{theorem}\label{theo:erasure_pattern}
Let $P\subset\{0,1\}^N$ be a puncturing pattern. Then $E[P] = P$ if and only if ${P}=\underline{0}$ (the all-zero vector) or $P$ is the union of a subset of rows of $\G_N$, that is, $\exists \mathcal{L} \subset \{0, \dots, N-1 \}$, ${P}=\bigvee\limits_{i \in \mathcal{L}} {G}^{(i)}_N$.
\end{theorem}

\begin{figure}
    \centering
    \includegraphics[scale=0.6]{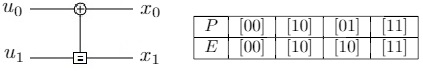}  
    \caption{Kernel bloc $\G_2$, and list of puncturing and erasure patterns}
    \label{fig:my_label2}
\end{figure}

\noindent\textit{Proof}: 
First we prove that if ${P}=\vee_{i \in \mathcal{L}} {G}^{(i)}_N$, then $E[P] = P$.  
Indeed, let ${\cal L}_{+} = \{0\leq i \leq N/2-1 \mid i\in{\cal L} \text{ or } N/2+i\in{\cal L} \}$, and ${\cal L}_{-} = \{0\leq i \leq N/2-1 \mid N/2+i\in{\cal L}\}$. Define ${ P}_{+} = \vee_{i \in \mathcal{L}_{+}} {G}^{(i)}_{N/2}$ and ${P}_{-} = \vee_{i \in \mathcal{L}_{-}} {G}^{(i)}_{N/2}$. Since ${G}_N^{(i)} \subset {G}_N^{(k)}$, for any  $k,i$, such that ${G}_N(k,i)=1$, it follows easily  that ${P}  = ({P}_{+}, P_{-})$. Then the equality ${E}[{P}] = {P}$ follows  by using induction arguments. 

\begin{figure}
    \centering
    \includegraphics[width=.6\linewidth]{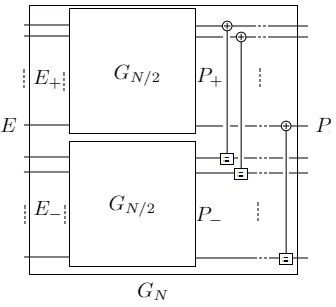}
    \caption{Recursive structure of a Polar code of length $N$}
    \label{fig:my_label4}
\end{figure}

To prove the direct implication, we prove that for any $P$, $E[P]$ is the  the union of a subset of rows of $\G_N$. The proof goes by induction on $n$. The statement can be easily verified for  $n=1$ ($N=2$), according to the list of erasure patterns provided in Fig.~\ref{fig:my_label2}. Let $N = 2^n$, with $n > 1$, and ${E} = {E} \lbrack {P} \rbrack$. Considering the recursive structure from Fig.~\ref{fig:my_label4}, the puncturing pattern ${ P}$ generates an erasure pattern on each of the $N/2$ kernel blocks on the right side of the figure. Putting them together, they produce two puncturing patterns on the top and bottom $\G_{N/2}$ blocks, indicated respectively by ${ P}_{+}$ and ${ P}_{-}$. 
Hence, ${ E}  = ({ E}_{+}, { E}_{-})$ is the concatenation of ${ E}_{+} = { E}[{ P}_{+}]$ and ${ E}_{-} = { E}[{ P}_{-}]$. By construction, and considering the possible erasure patterns for kernel blocks, it follows that ${ P}_{-}\subseteq { P}_{+}$, and therefore it can be easily verified that ${ E}_{-}\subseteq { E}_{+}$. By induction, ${ E}_{-} = \vee_{i \in \mathcal{L}_{-}} {G}^{(i)}_{N/2}$ and  ${ E}_{+} = \vee_{i \in \mathcal{L}_{+}} {G}^{(i)}_{N/2}$, with $\mathcal{L}_{-} \subseteq \mathcal{L}_{+} \subseteq \{0, \dots, N/2-1 \}$. It follows that ${ E} = \vee_{i \in \mathcal{L}} {G}^{(i)}_{N}$, where ${\cal L} = \mathcal{L}_{+} \cup \{N/2+\mathcal{L}_{-}\}$, and also using the fact that  ${G}^{(N/2+i)}_{N} = ({G}^{(i)}_{N/2}, {G}^{(i)}_{N/2})$, for any $0\leq i\leq N/2-1$.
\hfill $\square$

Note that the proof of the above theorem actually demonstrates that an erasure pattern is necessarily the the union of a subset of rows of $\G_N$. It also indicates how the erasure pattern ${ E}[{ P}]$ can be determined in practice: one has to propagate ${ P}$, from right to the left, throughout the recursive structure shown in Fig.~\ref{fig:my_label4}. At each propagation step, the pattern is propagated through a number of kernel blocks, according to the ${ P} \mapsto { E}[{ P}]$ rule shown in  Fig.~\ref{fig:my_label2}. In particular, this shows that the erasure pattern does not depend on the channel $W$, but only on the puncturing pattern $P$.

\begin{theorem}
If $P$ is a symmetric pattern, then $\Phi(P)=P$, \textit{i.e.} $P$ is the primitive pattern of its equivalence class.
\end{theorem}

\noindent\textit{Proof}: 
First, it can be proved that the bit-reversal permutation of any row of $G_N$ is also a row of $G_N$. Therefore, if $P$ is a symmetric pattern, then so is $B_N(P)$. Secondly, one can easily check that any row of $\G_N$, and therefore any symmetric pattern, is invariant by $\phi$. From the above observations, it follows that for any symmetric pattern $P$, one has: $\Phi(P)=B_N(\phi(B_N(P)))=B_N(B_N(P))=P$.
\hfill $\square$

Due to space limitations, the proof of the following theorem will be included in an extended version of the paper.

\begin{theorem}
Let $P$ be a puncturing pattern, $\bar{P} = B_N(P)$, $\bar{P}_0 = [\bar{P}(0),\dots, \bar{P}(N/2-1)]$ and 
$\bar{P}_1 = [\bar{P}(N/2),\dots, \bar{P}(N-1)]$. Then  $P$ is symmetric if and only if $B_{N/2}(\bar{P}_0)$ and $B_{N/2}(\bar{P}_1)$ are symmetric, and $\bar{P}_1\subseteq\bar{P}_0$. 
\end{theorem}


While the previous theorem indicates that symmetric patterns can be determined recursively (similarly to primitive patterns), the search-tree algorithm introduced in the next section is aimed at generating symmetric patterns, by exploiting their description as a union of a subset of rows of $\G_N$ (Theorem~\ref{theo:erasure_pattern}).

\subsection{Search-Tree Algorithm}\label{subsec:search_tree}

The \textit{order} $\lambda$ of a symmetric puncturing pattern $\mathcal{P}$ is defined as the minimum number of rows of $\G_N$ that generate $\mathcal{P}$. Hence $\lambda = \min\left\{ |\mathcal{L}| \mid P=\cup_{i \in \mathcal{L}} G_N^{(i)}\right\}$. It can be easily verified that a symmetric puncturing pattern $P$, of order $\lambda$ and weight $|P| = N_p$, is generated by a set $\mathcal{L}$ of  rows of $\G_N$, such that:
\begin{itemize}
\item $|{\cal L}| = \lambda$ and $\forall (i,j) \in \mathcal{L}^2, {G}_N^{(i)} \not\subseteq {G}_N^{(j)}$,
\item $|\vee_{i \in \mathcal{L}} {G}_N^{(i)}|=N_p$
\end{itemize}  
The search-tree algorithm proposed in this section generates symmetric puncturing patterns of order $\lambda\leq \lambda_{max}$, for some fixed $\lambda_{max}$ value. Note that for small $\lambda_{max}$ values, it might not be possible to get any value of $N_p$, thus resulting in a loss in flexibility with respect to the punctured code-length. However, the flexibility quickly increases with the considered maximum  order.

We denote by $L = (L(1),\dots, L(\lambda_{max}))$ a vector such that $0\leq L(\tau) \leq N$, for any  $\tau = 1,\dots,\lambda_{max}$. For $1\leq\lambda\leq\lambda_{max}$, we further denote $P_\lambda = \vee_{1\leq\tau\leq\lambda}G_N^{(L(\tau))}$. Hence, $L(1),\dots,L(\lambda)$ are the $\lambda$ indexes of the $\G_N$-rows contributing to the symmetric puncturing pattern $P_\lambda$. Since the rows of $\G_N$ are numbered from $0$ to $N-1$, we define $G_N^{(N)} = \underline{0}$, the all-zero vector. We also define $P_0 = \underline{0}$.

For given $N_p$ and $\lambda_{max}$ values, the search-tree algorithm works as follows:

\begin{enumerate}\setcounter{enumi}{-1}
\item Let $\lambda=1$ and $L(\tau) = N$, for any  $\tau = 1,\dots,\lambda_{max}$.

\item  Calculate the complementary weight values, defined by $W_{cp}(i) = |G_N^{(i)}|-|G_N^{(i)} \wedge P_{\lambda-1}|$, for $i = 0, \dots, N-1$.
\item Determine the largest index $i$ strictly less than $L(\lambda)$, such that:
\begin{equation}\label{eq:largest_index} 
\left\{ \begin{array}{cl}
0 < W_{cp}(i) \leq N_p - |P_\lambda|, & \text{if } \lambda < \lambda_{max} \\
W_{cp}(i) = N_p - |P_\lambda|, & \text{if } \lambda = \lambda_{max}
\end{array} \right. 
\end{equation}
If no index $i$ verifies Eq.~(\ref{eq:largest_index}), then set $\lambda=\lambda-1$ (upper tree level) and go to Step 4).  


\item Set $L(\lambda) = i$;\\
\begin{tabular}[t]{@{}p{10pt}@{}p{\linewidth-10pt}@{}}
-- & If $|P_{\lambda}| = |N_p|$, then store $P_{\lambda}$ (symmetric puncturing pattern of order $\lambda$ and weight $N_p$), and go back to Step 2) (continue the search on the same tree level);\\
-- & Otherwise, set $\lambda = \lambda+1$, then go back to Step 1) (continue search on the lower tree level).\\
\end{tabular}

\item If $\lambda = 0$ then all the puncturing patterns have been found and the process stops; otherwise, go to Step 1). 

\end{enumerate}
It can be noticed that the complexity of this algorithm is ${\cal O}(N^{\lambda_{max}})$, which allows generating symmetric patterns even for high $N$ values, assuming the $\lambda_{max}$ is relatively small.

\section{Numerical Results}\label{sec:num_results}

Throughout this section, we consider a Binary-Input Additive While Gaussian Noise (BI-AWGN) channel, with noise variance $\sigma^2$. For a given puncturing pattern $P$, the information pattern minimizing the $\text{WER}^{\text{\sc ga}}$ (Eq.~(\ref{eq:wer_approx})) depends on both $P$ and $\sigma^2$, and can be determined by DE, as explained in Section~\ref{sec:num2}. We denote this information pattern by $I(P, \sigma^2)$, and the corresponding word error rate by $\text{WER}^{\text{\sc ga}}(P, \sigma^2)$. Given a {\em target word error rate $\eta>0$}, we define the {\em noise (variance) threshold} $\sigma^2_{\text{th}} (P)$, as follows:
\begin{equation}
\sigma^2_{\text{th}} (P) = \sup \left\{\sigma^2 \mid \text{WER}^{\text{\sc ga}}(P, \sigma^2) \leq \eta\right\}
\end{equation}
It corresponds to the maximum $\sigma^2$ for which the punctured code provides a word error rate less than or equal to $\eta$, assuming an appropriate choice of the information pattern.


\subsection{Primitive vs. Symmetric Patterns}

\begin{table}
  \centering
  \caption{Number of primitive patterns for $N=64$}
  \label{tab:table2}
  \renewcommand{\arraystretch}{1.25}
  \begin{tabular}{|c|c|c|c|c|c|}
     \hline
    $N_p$ & 6 & 8 & 10 & 12 & 14\\
    \hline
    primitive & 381 & 2005 & 10599 & 42894 & 150502\\
    \hline
    symmetric & 156 & 605 & 2045 & 5913 & 14345\\
    \hline
    non-symmetric & 225 & 1600 & 8554 & 37281 & 136157\\
    \hline
  \end{tabular}
\end{table}

\begin{figure}[!t]
    \centering
    \includegraphics[width=\linewidth]{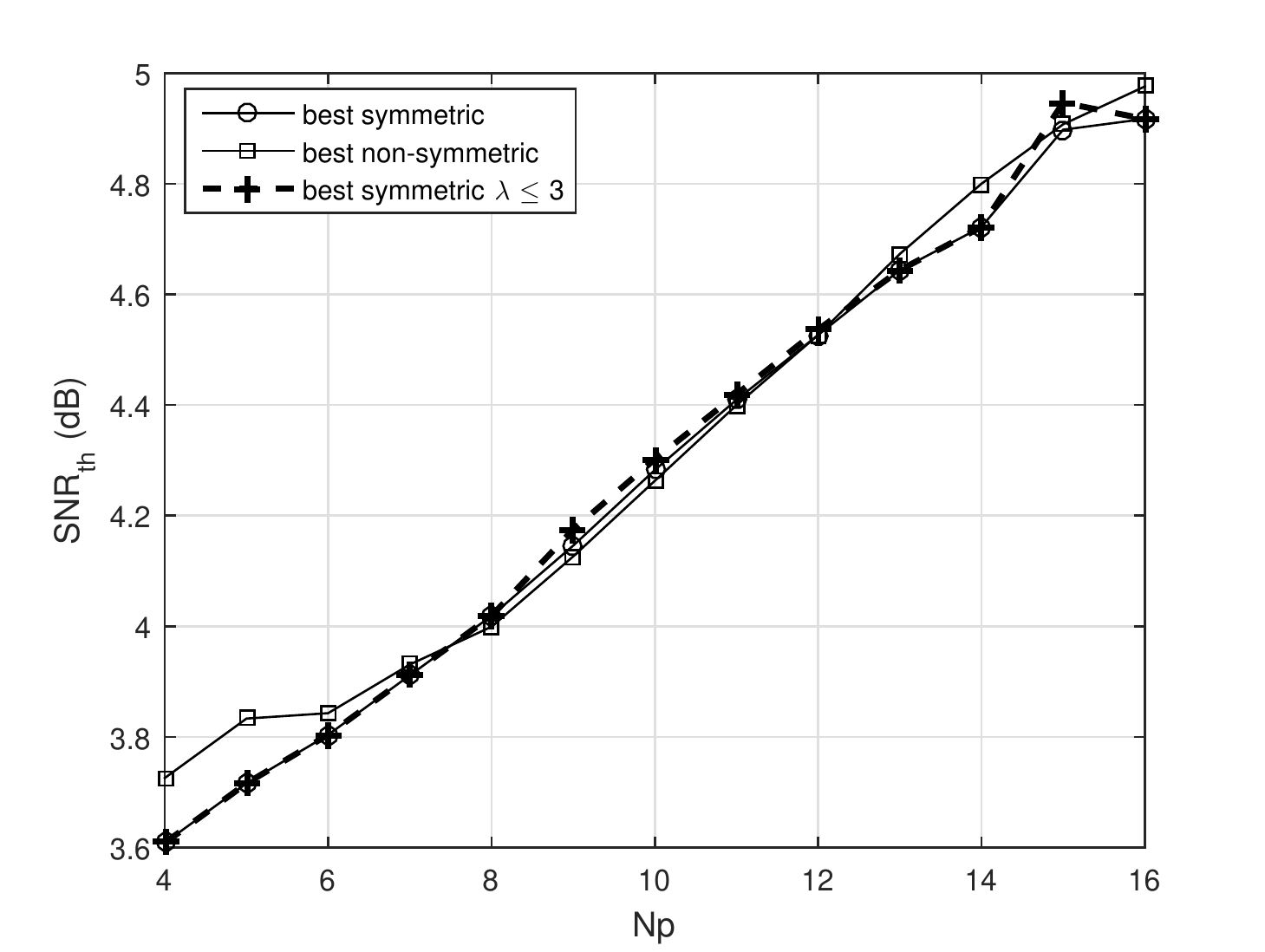}
    \caption{Comparisons of the best symmetric and non-symmetric patterns for $N=64$, $K=20$ and $\eta=10^{-4}$}
    \label{fig:my_label6}
\end{figure}

For $N=64$, we generate all the primitive puncturing patterns, by using the recursive construction method from Theorem~\ref{theo:primitive_reccursion}. We further class them in two categories, namely symmetric patterns and non-symetric ones.  The number of patterns in each category is provided in Table~\ref{tab:table2}. It can be seen  that the number of non-symmetric patterns increases much faster than the number of symmetric ones. For all the puncturing patterns, we use the DE technique\footnote{We use DE under the Gaussian Approximation hypothesis \cite{wu2014construction}} to determine the noise threshold for a target word error rate $\eta=10^{-4}$, assuming $K=20$ information bits. Fig.~\ref{fig:my_label6} shows the noise threshold (in dB) of the best puncturing patterns in each one of the two categories, for various $N_p$ values. We observe that the best symmetric and non-symmetric patterns have virtually the same noise threshold for any $N_p$ value. Therefore, one may restrict the search of good puncturing patterns to symmetric patterns only.  Moreover, Fig.~\ref{fig:my_label6} also shows that by considering only symmetric patterns of order $\lambda \leq 3$, the corresponding noise threshold values are very close to the optimal ones.  This justifies the optimization method considered in the next section, which consists in exploring only symmetric patterns of relatively low order.


\subsection{Symmetric vs. SotA Patterns}

\begin{figure}[!t]
    \centering
    \includegraphics[width=\linewidth]{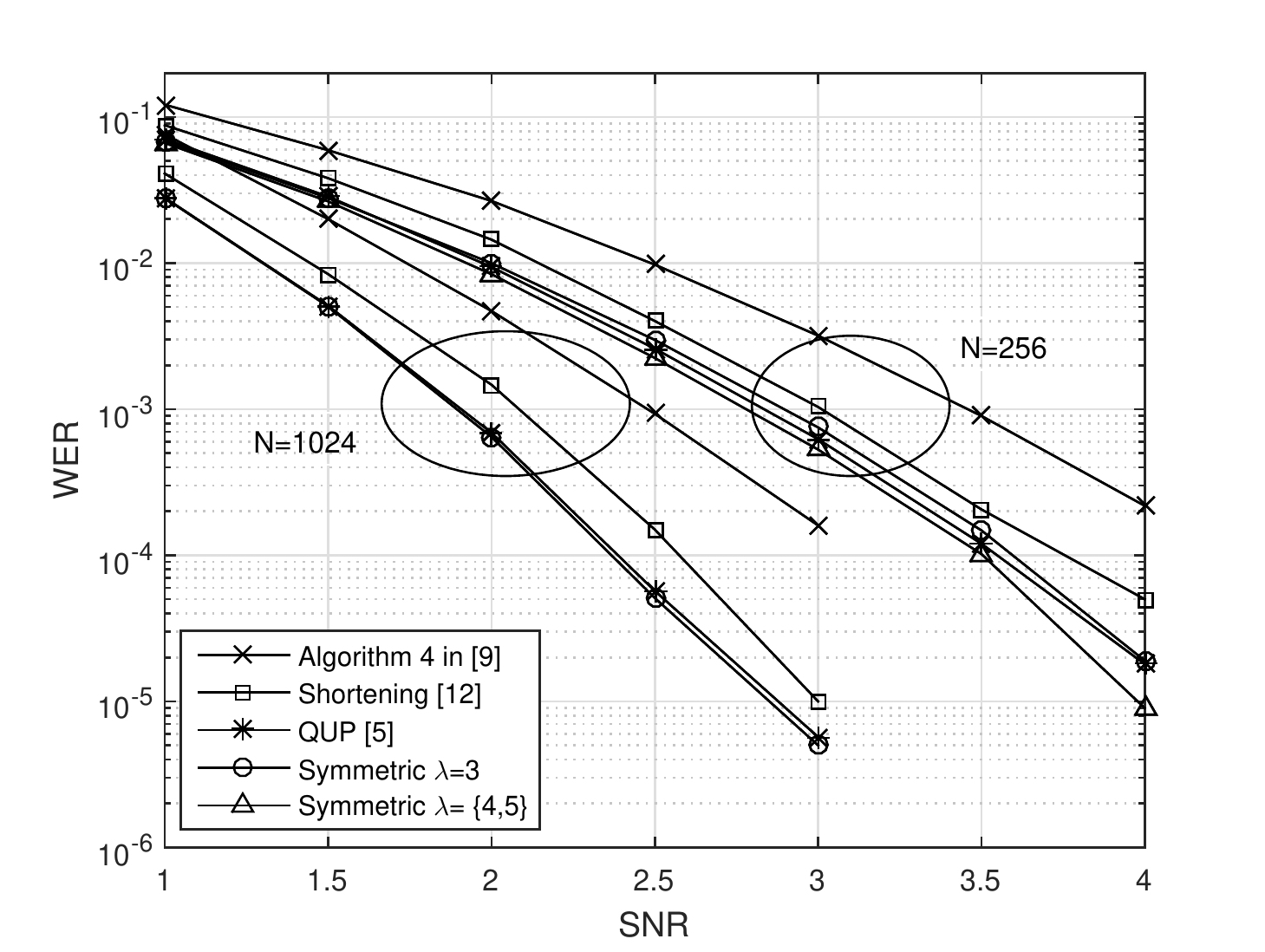}
    \caption{WER performance of rate-compatible codes with parameters $(N,K,N_p)=(256,64,85)$ and $(1024,256,336)$}
    \label{fig:my_label7}
\end{figure}

We consider the punctured codes with parameters $(N,K,N_p)=(256,64,85)$ and $(1024,256,336)$. For each $\sigma^2$ value, we rely on DE to find the symmetric puncturing pattern $P$ (and corresponding information pattern $I(P,\sigma^2)$) that minimizes $\text{WER}^\text{\sc ga}(P, \sigma^2)$. Due to the very large number of symmetric patterns, evaluation $\text{WER}^\text{\sc ga}(P, \sigma^2)$ for all of them is not feasible. Thus, we restrict our search to symmetric patterns of order $\lambda\leq\lambda_{max}$, by using the search-tree algorithm from Section~\ref{subsec:search_tree}. The WER performance of the punctured Polar code under SC decoding is then evaluated by Monte-Carlo simulation. Simulation results, are shown in Fig.~\ref{fig:my_label7}. For comparison purposes, we have also included the WER performance of rate compatible codes (with same $(N, N_p, K)$ parameters), obtained by using either puncturing \cite{niu2013beyond, zhang2014puncturing} or shortening \cite{wang2014novel} techniques from the state of the art. 

The number of symmetric patterns of order less than or equal to $\lambda_{max}$ is given in Table~\ref{tab:nb_sym_patterns_lmax}, for $\lambda_{max} = 3, 4, 5$. For $N=256$, we note that the best symmetric puncturing patterns of order less than or equal to $\lambda_{max}$, are the same for both $\lambda_{max} = 4$ and $\lambda_{max} = 5$. Hence, Fig.~\ref{fig:my_label7} contains only one corresponding plot, indicated by $\lambda_{max} = \{4,5\}$. This further confirms the observation from the previous section, concerning the relatively small order of the best puncturing patterns. 

Concerning the comparison with the state of the art, we observe that the proposed method outperforms the methods from \cite{zhang2014puncturing, wang2014novel}, and provides slightly better performance than the Quasi-uniform Puncturing (QUP) method \cite{niu2013beyond}, in the high SNR regime ($N=256$). It is worth noticing that, although the approach in \cite{niu2013beyond} is different from ours, it actually turns out that the QUP pattern is a symmetric pattern of order $\lambda = 4$ ($N = 256$) or $\lambda = 3$ ($N = 1024$).


\section{Conclusion}

In this paper, we studied equivalence classes of puncturing patterns and proposed a constructive solution to find a unique representative of each class, called primitive pattern. Primitive patterns have been further classified into symmetric patterns, equal to their erasure pattern, and non-symmetric ones. As the complexity of an exhaustive search through all the primitive patterns is impracticable even for relatively small $N$, we concentrated on the first category and indicated a method to generate symmetric patterns, by using a search-tree algorithm that exploits their characterization as union of rows of the $\G_N$  matrix. Simulations showed that the proposed method performs better than the main methods of the state-of-the-art.


\begin{table}
  \centering
  \caption{Number of symmetric patterns of order  $\leq \lambda_{max}$}
  \label{tab:nb_sym_patterns_lmax}
  \renewcommand{\arraystretch}{1.25}
  \begin{tabular}{|c||c|c|c|}
     \hline
    $\lambda_{max}$ & 3 & 4 & 5 \\
    \hline\hline
    $N=256$  & $2940$ & $3.5 \cdot 10^{5}$ & $13.5 \cdot 10^{6}$ \\
    \hline
    $N=1024$ & $3 \cdot 10^5$ & --  & -- \\
    \hline
  \end{tabular}
\end{table}


\section*{Acknowledgment}

The research leading to these results received funding from the European Commission H2020 Programme, under grant agreement 671650 (mmMagic Project).

\bibliographystyle{IEEEtran}
\bibliography{biblio_puncturing}

\end{document}